\documentclass{paper}
\usepackage{amsfonts,amsmath,latexsym,amscd,graphicx,amssymb}

\begin{document}

\title{Peakons arising as particle paths beneath small-amplitude water waves}

%\label{firstpage}

\author{\normalsize Delia IONESCU-KRUSE\\
\normalsize Institute of Mathematics of
the Romanian Academy,\\
\normalsize P.O. Box 1-764, RO-014700, Bucharest,
 Romania\\
\normalsize E-mail: Delia.Ionescu@imar.ro\\[10pt]}

 \date{}

\maketitle

\newcommand{\be}{\begin{equation}}
\newcommand{\ee}{\end{equation}}
\newcommand{\ba}{\begin{eqnarray}}
\newcommand{\ea}{\end{eqnarray}}
\newcommand{\pa}{\partial}
\newcommand{\f}{\frac}
\newcommand{\st}{\stackrel}
\newcommand{\s}{\sqrt}

%\begin{history}
%\received{(Day Month Year)} \revised{(Day Month Year)}
%\accepted{(Day Month Year)}
%\comby{(xxxxxxxxx)}
%\end{history}

\begin{abstract}
%We present one possible trajectory of the particles  beneath
%small-amplitude gravity waves in constant vorticity water flow
%with a flat bottom.

We present a new kind of  particle path in constant vorticity
water of finite depth, within the framework of small-amplitude
waves.
\end{abstract}

%\keywords{Keyword1; keyword2; keyword3.}

%\ccode{2000 Mathematics Subject Classification: }

\section{Introduction}

A \textit{peakon}  is a soliton with discontinuous first
derivative \cite{holm}.
%The wave profile is shaped like the graph
%of the function $\exp(-| x |)$.
The concept was introduced in 1993
by Camassa and Holm in the paper \cite{ch}, where they derived the
CH shallow water equation \begin{equation} u_t+2\kappa
u_x+3uu_x-u_{txx}=2u_xu_{xx}+uu_{xxx},  \quad
 \quad \textrm{(CH)}
\label{ch}\end{equation} $(x,t)\in\mathbf{R}\times (0,\infty)$,
$\kappa$ being a real constant. Alternative derivations of CH
equation are provided  in the papers \cite{johnson}, \cite{c&l},
\cite{io}. The peakons arise as solution of this equation for
$\kappa$=0. The CH peakons are given by \be u(x,t)=c
\exp(-|x-ct|), \quad c\in\mathbf{R}. \label{peak}\ee Since peakon
solutions are only piecewise differentiable, they must be
interpreted in a suitable weak sense. The derivative \be
u_x=-c\textrm{ sgn }(x-ct)\exp(-|x-ct|) \ee has a jump
discontinuity at the peak. The second derivative $u_{xx}$ must be
taken in the sense of distributions and will contain a Dirac delta
function \be u_{xx}= c \exp(-|x-ct|)-2c\delta(x-ct)\ee The
function $m$ is defined by \be m(x,t):=u-u_{xx}=
2c\delta(x-ct).\label{momentum}\ee Physically $m$ has the
interpretation of momentum \cite{ch}, \cite{homara}.

\noindent  The peakon (\ref{peak}) has amplitude $c$ and travels
at speed $c$. At $x=ct$ the momentum (\ref{momentum}) blows up at
$+\infty$.

\noindent  A small perturbation of a CH peakon yields another one
which remains close to some translate of the initial one at all
later times. In this sense the CH peakons are orbitally stable
\cite{cs1}.
%By taking a linear combination of peakons one obtains
%what is called a multipeakon solution.
Of particular interest is the description of peakon dynamics in
terms of a system of completely integrable Hamiltonian equations
for the locations of the peaks of the solution. Thus, each peakon
solution can be associated with a mechanical system of moving
particles. Being solitons, they retain their shape and speed after
interacting with other peakons \cite{BSS2}. The peakon interaction
plays an important role in the general dynamics of the solutions
to the equation (see the discussion in \cite{holden}) and provided
the framework for the construction of global weak solutions both
in the conservative case  \cite{bc1} as well as in the dissipative
case  \cite{bc2}.
 One of the main interests in CH equation was that, in contrast to
other standard shallow water equations, as for example the KdV
equation, it models breaking waves: smooth solutions that develop
singularities in finite time, the solution being bounded but its
slope becoming unbounded. This fact was already noted in \cite{ch}
and subsequently proved in \cite{ce}.

Another completely integrable CH-type equation which has peakon
solutions  \cite{dhh} is the Degasperis-Procesi equation \cite{dp}
\begin{equation} u_t+4uu_x-u_{txx}=3u_xu_{xx}+uu_{xxx}, \quad \quad \textrm{(DP)}
\label{dp}\end{equation} $(x,t)\in\mathbf{R}\times (0,\infty)$.
The DP equation possesses not only peaked solitons (\ref{peak})
but also discontinuous solitons, so-called \textit{shock-peakons}
\cite{lund} of the form \be u(x,t)=c\exp(-|x|)-\f 1{t+k}\textrm{
sgn }(x)\exp(-|x|), \quad k>0 \label{shock} \ee At the peak they
have a finite jump in the function $u$ itself. The shock-peakon
solutions must be interpreted in a proper weak formulation. The
derivative $u_x$ will contain $\delta$ and the function
$m:=u-u_{xx}$ will be a linear combination of $\delta$ and
$\delta'$ distributions. It is not known to the author if the
function $m$ can be in this case interpreted as momentum.  We
point out that the  CH equation with $\kappa=0$, $\kappa\neq 0$,
is a geodesic equation on the diffeomorphism group of the circle
\cite{ck},
 respectively on the Bott-Virasoro group
\cite{misiolek},  \cite{ckkt}, while the DP equation is a
non-metric equation  \cite{ek}.
\\
The shock-peakon (\ref{shock})  moves at constant speed $c$ (in
particular, does not move if $c=0$ \cite{lund}) which is equal to
the average amplitude at the jump. The shock "dissipates away"
like $1/t$ as $t\rightarrow +\infty$.

\noindent The peakons of the DP equation are also true solitons
that interact via elastic collisions under the DP dynamics
\cite{lunds}, and are also  orbitally stable \cite{lin}.

In what follows we will see that in the study of  particle motion
beneath  small-amplitude water waves in constant vorticity flows a
peakon trajectory comes up. This solution contains
arctanh$(\cdot)$ function, having a vertical asymptote in the
positive direction.

\section{Particle path
 beneath  small-amplitude water waves in constant
vorticity flows}

We consider  two-dimensional gravity waves on constant vorticity
water of finite depth. They  are described, in non-dimensional
scaled variables, by the following boundary value problem (see,
for example \cite{johnson-carte}):
\begin{equation}
\begin{array}{cc}
u_t+\epsilon(uu_x+vu_z)=- p_x&\\  \delta^2[v_t+\epsilon(uv_x+vv_z)]=- p_z&\\
 u_x+v_z=0&\\
  u_z=\delta^2v_x+\f{\sqrt{gh_0}}{g}\omega_0&\\
  v=\eta_t+\epsilon u\eta_x  \, & \textrm{ on }\,
z=1+\epsilon\eta(x,t)\\
p=\eta \, & \textrm{ on }\,
z=1+\epsilon\eta(x,t)\\
 v=0 \, &
\textrm { on } z=0
 \end{array}
\label{e+bc1''} \end{equation}
  where $(x,z)$ are the space coordinates, $(u(x,z,t), v(x,z,t))$ is
the velocity field of the water, $p(x,z,t)$ denotes the pressure,
$g$ is the constant gravitational acceleration in the negative $z$
direction, $\omega_0$ being the constant  vorticity. We have
introduced the amplitude parameter $\epsilon=\f a{h_0}$ and the
shallowness parameter $\delta=\f {h_0}{\lambda}$, with $a$ the
amplitude of the wave and $\lambda$ the wavelength. $h_0>0$ is the
undisturbed depth of the fluid and $z=1+\epsilon \eta(x,t)$
represent the free upper surface of the fluid in non-dimensional
scaled variables. The existence of solutions of large and small
amplitude was recently proved in \cite{cs2} where it is also shown
that linearization provides an accurate approximation for waves of
small amplitude.

 By letting $\epsilon\rightarrow
0$, $\delta$  being fixed, we obtain a linear approximation of
 our problem, that is,
% we
%get the following linear system
\begin{equation}
\begin{array}{cc}
u_t+p_x=0&\\ \delta^2v_t+ p_z=0&\\
 u_x+v_z=0&\\
 u_z=\delta^2v_x+\f{\sqrt{gh_0}}{g}\omega_0&\\
v=\eta_t  \, & \textrm{ on }\,
z=1\\
  p=\eta \, & \textrm{ on }\,
z=1\\
 v=0 \, &
\textrm { on } z=0
\end{array}
\label{small} \end{equation}
The system (\ref{small}) has the
solution \be
 \begin{array}{llll}
 \eta(x,t)=\cos(2\pi(x-ct))\\
u(x,z,t)=\f{2\pi\delta c}{\sinh (2\pi\delta)}\cosh (2\pi\delta
z)\cos(2\pi(x-ct))+\f{\omega_0\sqrt{gh_0}}{g}z+c_0\\
v(x,z,t)=\f{2\pi c}{\sinh(2\pi\delta)}\sinh (2\pi\delta
z)\sin(2\pi(x-ct))\\
 p(x,z,t)=\f{2\pi\delta
c^2}{\sinh(2\pi\delta)}\cosh(2\pi\delta z)\cos(2\pi(x-ct))
\end{array}\label{solrot0}\ee with
   the non-dimensional speed of the linear wave given by \be
c^2=\f{\tanh(2\pi\delta)}{2\pi\delta}\label{c}\ee

Let $\left(x(t), z(t)\right)$ be the path of a particle in the
fluid domain, with location $\left(x(0), z(0)\right):=(x_0,z_0)$
at time $t=0$. The motion of the particles  below the
 small-amplitude gravity water
waves given by (\ref{solrot0}), is described by the following
system of differential equations
  \be\left\{\begin{array}{ll}
 \f{dx}{dt}=u(x,z,t)=\f{2\pi\delta c}{\sinh (2\pi\delta)}\cosh (2\pi\delta
z)\cos(2\pi(x-ct))+\f{\omega_0\sqrt{gh_0}}{g}z+c_0\\
\\
 \f{dz}{dt}=v(x,z,t)=\f{2\pi c}{\sinh(2\pi\delta)}\sinh (2\pi\delta
z)\sin(2\pi(x-ct))
 \end{array}\right.\label{diff2}\ee
 Notice that the constant $c_0$ is the average of the horizontal
fluid
 velocity on the bottom over any horizontal
 segment of length 1, that is,
 \be
 c_0=\f 1 {1}\int_{x}^{x+1}u(s,0,t)ds
 \ee
This is accordance with Stokes' definition of the wave speed for
irrotational flows (see the discussion in \cite{cs3}).

To study the exact solution of the system (\ref{diff2}) it is
 more convenient to rewrite it in the following moving frame
 \be
 X=2\pi(x-ct),\quad  Z=2\pi\delta z \label{frame}
 \ee
This transformation yields \be\left\{\begin{array}{ll}
 \f{dX}{dt}=\f{4\pi^2\delta c}{\sinh(2\pi\delta)}\cosh(Z)\cos(X)+
\f{\omega_0\sqrt{gh_0}}{g\delta}Z +
 2\pi(c_0-c)\\
 \\
 \f{dZ}{dt}=\f{4\pi^2\delta c}{\sinh(2\pi\delta)} \sinh(Z)\sin(X)
 \end{array}\right.\label{diff3}\ee
We denote by \be A:= \f{4\pi^2\delta
 c}{\sinh(2\pi\delta)}\quad \textrm{ and } \quad
 \Omega_0:=\f{\omega_0\sqrt{gh_0}}{g\delta}
\label{a2}\ee With the notations (\ref{a2}), the system
(\ref{diff3}) becomes: \be\left\{\begin{array}{ll}
 \f{d X}{dt}=A\cosh(Z)\cos(X)+\Omega_0Z+2\pi(c_0-c)\\
 \\
 \f{d Z}{dt}=A  \sinh(Z)\sin(X)
 \end{array}\right.\label{30'}\ee
We write the second equation of this system in the form \be
\f{dZ}{\sinh(Z)}=A\sin X(t)\,dt \label{44}\ee Integrating, we get
\be \log\left[\tanh\left(\f Z{2}\right)\right]=\int A \sin
X(t)\,dt \ee If \be \int A \sin X(t)\,dt<0 \label{int}\ee then \be
Z(t)=2\textrm{ arctanh }\left[\exp\left(\int A \sin
X(t)\,dt\right)\right] \label{42}\ee Taking into account the
formula: \be \cosh(2x)=\f{1+\tanh^2(x)}{1-\tanh^2 (x)},
\label{45}\ee and the expression (\ref{42}) of $Z(t)$, the first
equation of the system (\ref{30'}) becomes
 \be \f{d
X}{dt}=A\f{1+w^2}{1-w^2}\cos (X)+2 \Omega_0 \textrm{ arctanh
}(w)+2\pi(c_0-c)\label{35'}\ee where we have denoted by \be
w=w(t):=\exp\left(\int A \sin X(t)\,dt\right) \label{31}\ee With
(\ref{int}) in view, we have \be 0<w<1\label{w} \ee
 From (\ref{31}) we get \be
A\sin X(t)=\f 1{w(t)}\f{d w}{dt} \label{32}\ee Differentiating
with respect to $t$ this relation, we obtain
 \be A\cos (X)\f{dX}{dt}=\f 1{w^2}\left[\f{d^2w}{dt^2}w-\left(\f{dw}{dt}
 \right)^2\right]\label{33}
\ee From (\ref{32}) we have furthermore \be A^2\cos^2 (X)=A^2-\f
1{w^2}\left(\f{d w}{dt}\right)^2\label{34} \ee Thus, taking into
account (\ref{33}), (\ref{34}), the equation (\ref{35'}) becomes
\begin{eqnarray} &&\f{d^2w}{dt^2}+\f{2w}{1-w^2}\left(\f{dw}{dt}
\right)^2-A^2 w\f{1+w^2}{1-w^2}-\nonumber\\
 &&\hspace{1cm}- \sqrt{A^2w^2-\left(\f{dw}{dt}
 \right)^2}\Big[2\Omega_0\textrm{ arctanh }(w)+2\pi(c_0-c)\Big]=0\label{36'}
\end{eqnarray}
 We make
the following substitution
 \be \xi^2(w):=A^2w^2-\left(\f{dw}{dt}\right)^2 \label{37}\ee
 Differentiating with respect to $t$ this relation, we get
 \be
\xi\f{d\xi}{dw}=A^2w-\f{d^2w}{dt^2}
 \label{38}\ee
We replace (\ref{37}), (\ref{38}) into the equation (\ref{36'})
and we obtain  the equation
 \be
\xi\f{d\xi}{dw}+\f{2w}{1-w^2}\xi^2+\Big[2\Omega_0\textrm{ arctanh
}(w)+2\pi(c_0-c)\Big]\xi=0 \label{39c}\ee A solution of the
equation (\ref{39c}) is \be \xi=0\ee which, in view of (\ref{37})
and (\ref{32}) implies \be \sin X(t)=\pm 1 \ee Therefore, from
(\ref{42}) with the condition (\ref{int}),  and further from
(\ref{frame}), a solution of the system (\ref{diff2})  is \be
\begin{array}{ll}
 x(t)=ct+k_1\\
 \\
 z(t)=\f 1{\pi\delta}\textrm{ arctanh }\left[\exp\left(- A |t|\right)\right]
  \end{array} \label{sol0}\ee
$k_1$ being a constant.
  We observe that
\be \lim_{t\rightarrow 0}x(t)=k_1,\quad
\lim_{\substack{t\rightarrow 0\\t>0
}}z(t)=\lim_{\substack{t\rightarrow 0\\t<0 }}z(t)=+\infty \ee and
\be \lim_{t\rightarrow \pm\infty}x(t)=\pm\infty,\quad
\lim_{t\rightarrow \pm\infty}z(t)=0 \ee Therefore,  $x=k_1$ will
be a vertical asymptote and $z=0$ will be a horizontal asymptote
for  the curve (\ref{sol0}).
  The graph of the parametric curve (\ref{sol0}) is drawn  in
  figure 1.

\vspace{0.6cm}

\hspace{0cm}\scalebox{0.50}{\includegraphics{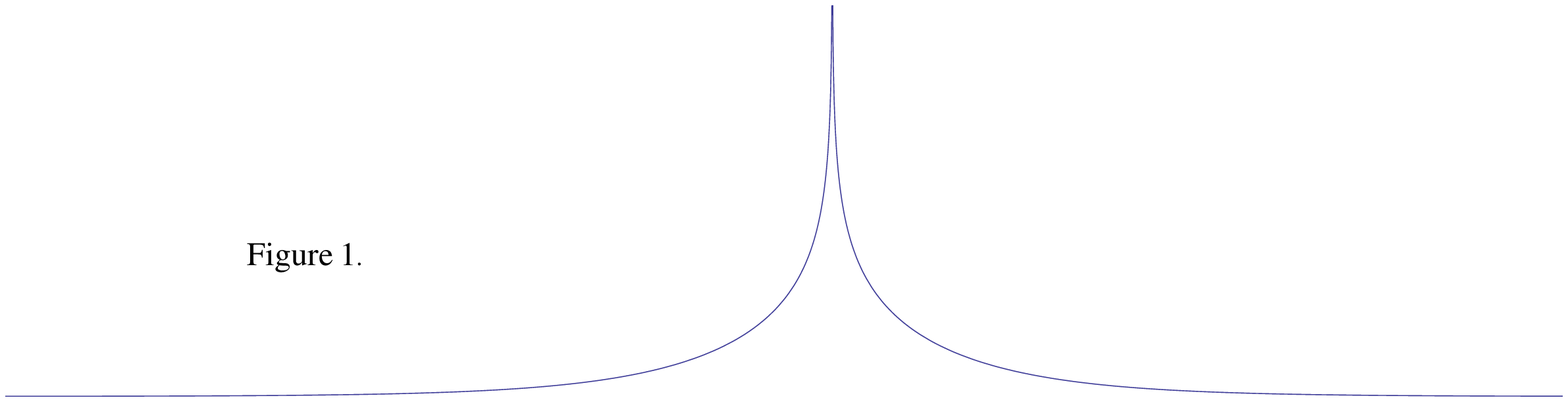}}
\\

Notice that within the setting of irrotational flows with no
underlying current (see \cite{c}) there are no such paths but in
the context of irrotational flows with an underlying (uniform)
current, the possibility of such paths was already noticed: see
\cite{cs3} for the exact solutions,  where this shape can be
thought of as a limiting case of the situation depicted in Figure
4.4 (ii), as well as  \cite{io2} for the linearized problem, where
somewhat similar particle paths are encountered.

\end{document}